\documentclass[conference]{IEEEtran}
\IEEEoverridecommandlockouts
\usepackage{xurl}
\usepackage{natbib}
\usepackage{amsmath,amssymb,amsfonts}
\usepackage{algorithmic}
\usepackage{graphicx}
\usepackage{textcomp}
\usepackage{xcolor}
\usepackage{xspace}
\usepackage{tikz}
\usepackage{float}
\usetikzlibrary{shapes,arrows,backgrounds,fit,positioning,calc,intersections,patterns.meta}
\def\BibTeX{{\rm B\kern-.05em{\sc i\kern-.025em b}\kern-.08em
    T\kern-.1667em\lower.7ex\hbox{E}\kern-.125emX}}
\newcommand{\aquillm}{AquiLLM\xspace}
\begin{document}

\title{\textit{AquiLLM}: a RAG Tool for Capturing Tacit Knowledge in Research Groups

}

\author{\IEEEauthorblockN{Chandler Campbell}
\IEEEauthorblockA{\textit{Department of Computer Science} \\
\textit{Southern Oregon University}\\
Ashland OR, USA \\
campbellr@sou.edu}
\and
\IEEEauthorblockN{Bernie Boscoe}
\IEEEauthorblockA{\textit{Department of Computer Science} \\
\textit{Southern Oregon University}\\
Ashland OR, USA \\
boscoeb@sou.edu}
\and
\IEEEauthorblockN{Tuan Do}
\IEEEauthorblockA{\textit{Physics \& Astronomy Department} \\
\textit{University of California, Los Angeles}\\
Los Angeles CA, USA \\
tdo@astro.ucla.edu}

}

\maketitle

\begin{abstract}

Research groups face persistent challenges in capturing, storing, and retrieving knowledge that is distributed across team members. Although structured data intended for analysis and publication is often well managed, much of a group's collective knowledge remains informal, fragmented, or undocumented—often passed down orally through meetings, mentoring, and day-to-day collaboration. This includes private resources such as emails, meeting notes, training materials, and ad hoc documentation. Together, these reflect the group's tacit knowledge—the informal, experience-based expertise that underlies much of their work. Accessing this knowledge can be difficult, requiring significant time and insider understanding. Retrieval-augmented generation (RAG) systems offer promising solutions by enabling users to query and generate responses grounded in relevant source material. However, most current RAG-LLM systems are oriented toward public documents and overlook the privacy concerns of internal research materials. We introduce AquiLLM (pronounced ah-quill-em), a lightweight, modular RAG system designed to meet the needs of research groups. AquiLLM supports varied document types and configurable privacy settings, enabling more effective access to both formal and informal knowledge within scholarly groups.

\end{abstract}


\section{Introduction}
%% BB: why does it say Sloan Foundation on the bottom left of first page

\subsection{Background and Motivation}
Research groups maintain extensive collections of information assets critical to their scientific endeavors. At the most formal level, they preserve published literature such as peer-reviewed journal articles, conference proceedings, and reference textbooks that document established knowledge in their field. These publications provide the theoretical foundations and experimental precedents that inform current research directions.

Beyond formal publications, research groups accumulate substantial technical documentation including laboratory protocols, instrument manuals, computational methods, code and code repositories \cite{szalay_designing_2000}. This information captures the precise methodological details required for experimental replication and data analysis \cite{peng_reproducible_2011}. Groups often develop specialized tools, algorithms, and procedures customized to their specific research questions, and in the process generate documentation that rarely appears in formal publications, but is essential for daily operations \cite{paine_who_2017}.

Perhaps most valuable is the informal knowledge generated through the research process itself. Laboratory notebooks document experimental procedures and outcomes, including failed attempts and unexpected observations that may never reach publication \cite{randles_using_2017}\cite{rule_exploration_nodate}. Meeting notes capture evolving hypotheses, proposed solutions to technical challenges, and collaborative brainstorming sessions. Email exchanges between collaborators often contain critical insights or clarifications about experimental design or data interpretation. This informal knowledge represents the intellectual "glue" that connects formal publications with the messy reality of scientific discovery and provides context for understanding why certain approaches were pursued or abandoned \cite{kernan_freire_tacit_2023}. For groups writing code as part of their research practice, this category may also include software tools, custom scripts, or computational pipelines, which function as key artifacts of research practice \cite{kitchin_codespace_2011}.

These artifacts contain tacit knowledge: the informal, experience-based understanding shared across a team \cite{polanyi_tacit_1967}\cite{collins_tacit_2013}. Such knowledge is often passed down orally through mentoring, collaborative meetings, or day-to-day problem-solving. It reflects the nuanced reasoning behind methodological decisions, the context surrounding past projects, and the practical expertise that supports ongoing research.

However, accessing this knowledge can be time-consuming and difficult. Traditional search tools depend on exact keyword matches and do not handle variation in terminology or document formats. Moreover, relevant information may be scattered across different storage systems, requiring familiarity not just with the scientific content but also with the group’s internal communication patterns and data conventions \cite{wallis_distribution_2012}.
%% there is no mention of code as its own thing as we discussed-- if so, add a sentence about code as some kind of artifact too, and cite KITCHIN CODE/SPACE software and everyday life which is different than the KITCHIN 2016

Finding specific information within a research group's knowledge corpus presents a significant challenge. Traditional approaches such as using Ctrl + F or command-line tools such as \texttt{grep} rely on exact text matching, which requires researchers to know precisely what terminology to search for. This becomes problematic when concepts are described using different vocabulary between documents or when researchers are unfamiliar with the specific terminology used in historical group documents. These methods cannot account for semantic relationships or conceptual similarities between pieces of text that lack lexical overlap \cite{karpukhin_dense_2020}.

The fragmentation of information across multiple types of documents further complicates retrieval efforts. A complete answer to a research question might require synthesizing information from formal publications, experimental notes, email exchanges, and meeting minutes. Without a unified search system, researchers must manually review numerous documents, making educated guesses about where relevant information might be located. This process is particularly time-consuming for new group members who lack institutional knowledge to narrow their search efficiently.

Perhaps the most challenging task is resolving inconsistencies when documents contain conflicting information. Research evolves over time, with hypotheses being refined, methodologies improved, and conclusions reconsidered. Older documents may contain outdated information that contradicts more recent findings. Traditional search methods provide no mechanism for identifying or reconciling these contradictions, leaving researchers to determine which source is most authoritative based on contextual cues such as document dates, authorship, or subsequent citations.

Recent advances in retrieval-augmented generation (RAG) systems, which combine information retrieval with large language models (LLMs), offer new possibilities for surfacing relevant content from large document corpora \cite{lewis_retrieval-augmented_2020}. Yet most current RAG-LLM applications focus on public-facing resources such as published literature or open-source documentation. They are not well suited for research group settings where internal materials are diverse, sensitive, and often semi-structured or informal.

We introduce AquiLLM, a lightweight, modular RAG system designed specifically for research groups. AquiLLM integrates varied internal document types, respects privacy configurations, and supports researchers in retrieving both formal and informal knowledge efficiently. By making tacit knowledge more accessible, AquiLLM aims to improve collaboration, onboarding, and institutional memory within scholarly teams.
%% this is where BB stopped adding citations and edits
\begin{figure*}
    \includegraphics[width=\textwidth]{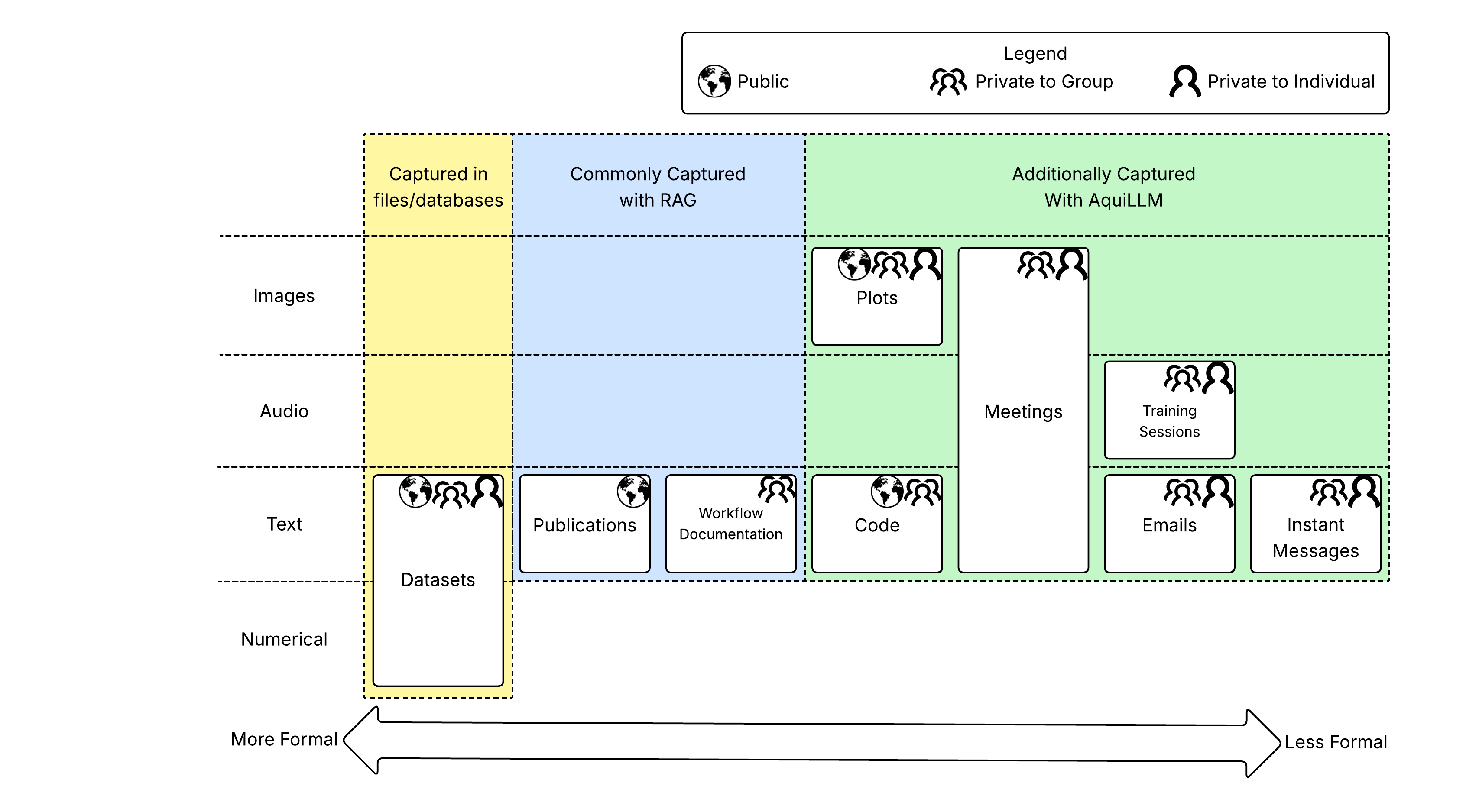}
    \caption{The types of data, modalities, and required privacy levels of information captured by traditional files and databases, common RAG systems, and \aquillm.}
    \label{fig:datatypes}
\end{figure*}
\begin{figure*}
   \centering
    \includegraphics[width=0.7\textwidth]{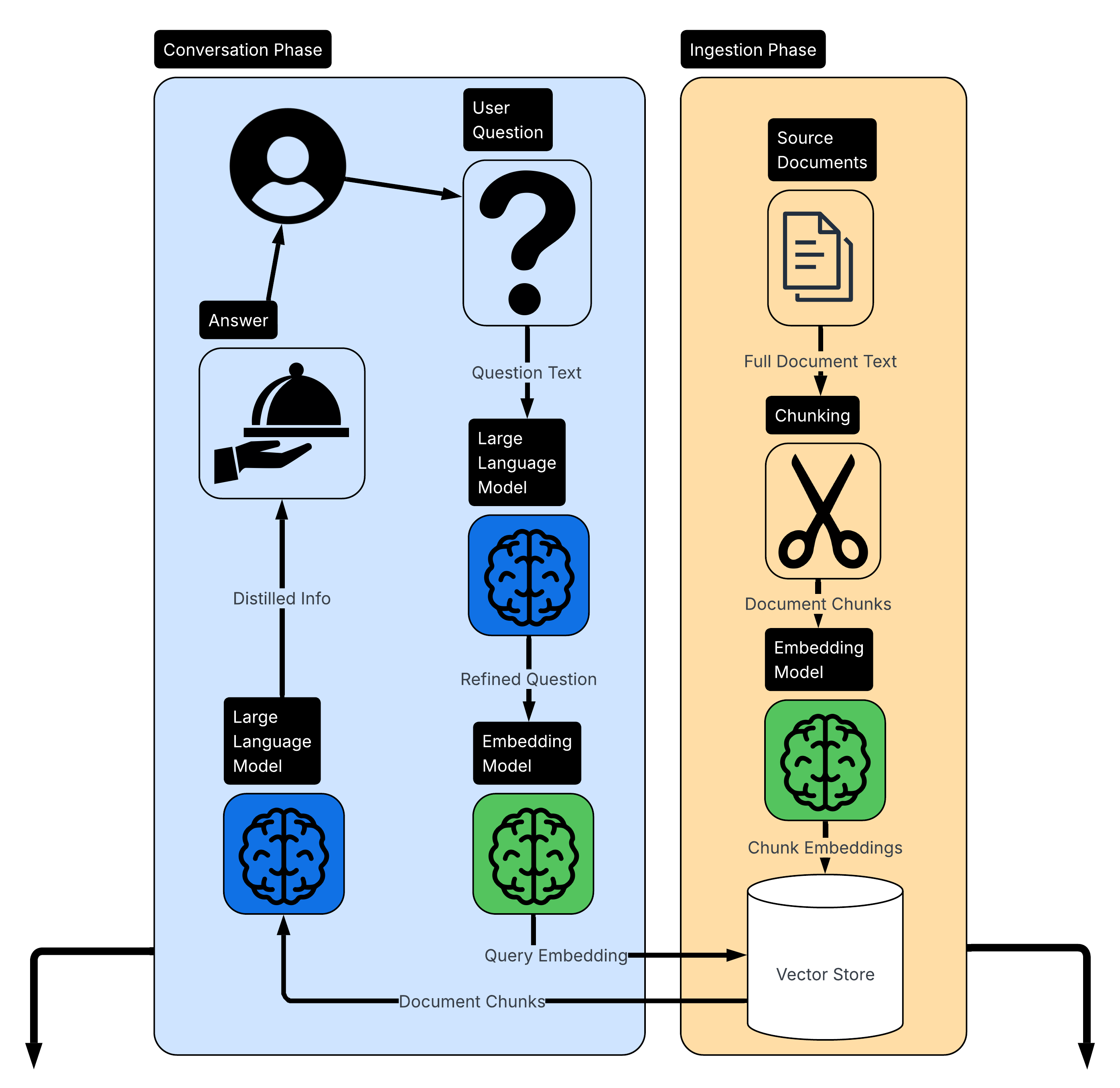} \\

   \begin{tabular}{cc}

      \includegraphics[width=0.4\textwidth]{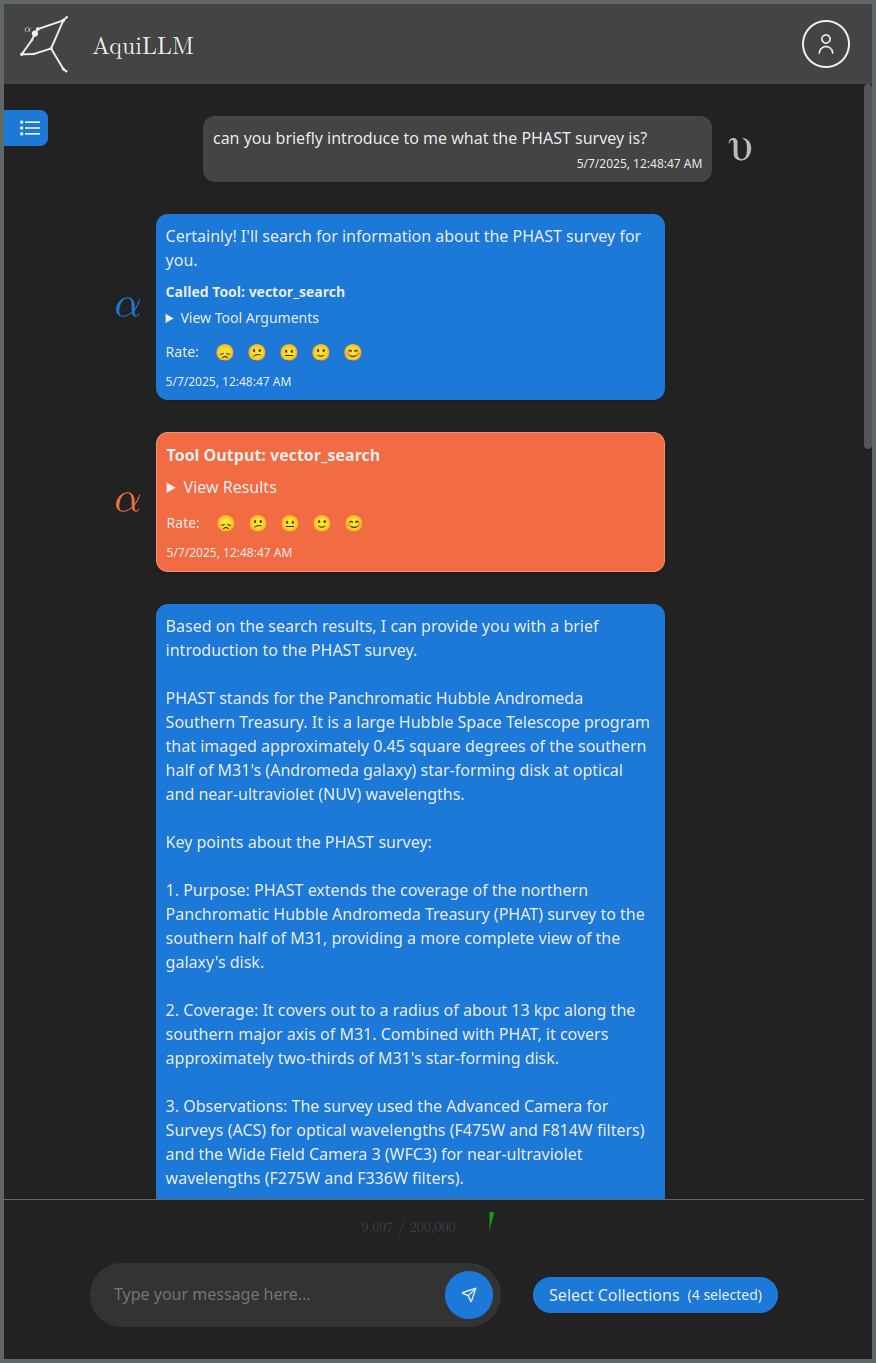} &
      \includegraphics[width=0.4\textwidth]{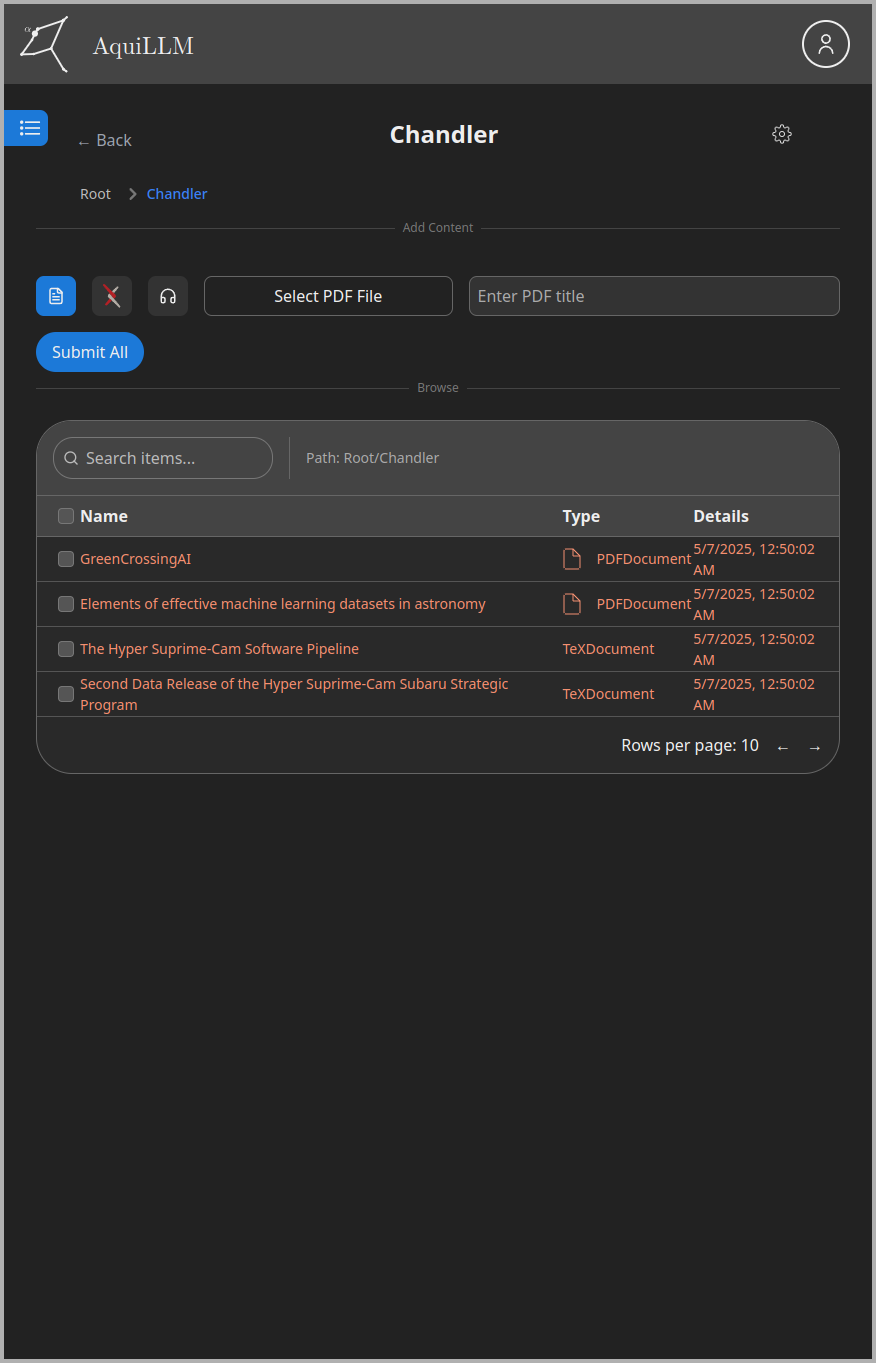}\\
   \end{tabular}
   
   \caption{The flow of information through \aquillm.}
   \label{fig:flowchart}
   
\end{figure*}

\subsection{The Case for RAG}
A retrieval-augmented generation (RAG) system tailored for research groups addresses these information retrieval challenges. By combining vector-based semantic search with the abilities of large language models to comb through large amounts of text, such a system can help researchers find answers from corpora which are otherwise difficult to navigate. Researchers can pose natural language questions without knowing exact terminology, and the system can identify relevant information across documents that use different vocabulary to describe the same concepts \cite{karpukhin_dense_2020}. This semantic understanding dramatically reduces the time spent searching, particularly for newer group members unfamiliar with group-specific terminology and document organization.

The unified indexing approach of a research-oriented RAG system overcomes document fragmentation by creating a comprehensive knowledge base from diverse source materials. When queried, the system can synthesize information scattered across formal publications, experimental notes, meeting minutes, and communications into coherent, contextual responses. This capability is extremely valuable for answering complex questions that require integrating information from multiple sources, a task that otherwise demands extensive manual review and mental synthesis by researchers who already possess significant domain expertise.

When confronted with conflicting information, the LLM embedded in a RAG system can provide the user with temporal context and highlight discrepancies between sources. By searching a large corpus, the system can present important historical context, helping researchers understand how the group's thinking has evolved over time. By presenting these information conflicts the system enables researchers to make informed judgments about which information to trust while maintaining institutional memory of alternative perspectives or approaches. This capability is particularly valuable for preserving institutional memory during personnel transitions, ensuring that hard-won insights are not lost when experienced members depart.

Finally, a RAG system that incorporates a \textit{shared} document store can enhance research collaboration by hosting a centralized knowledge base accessible to all team members. The system can serve as an information repository and a hub for collaboration, allowing researchers to build upon each other's work more effectively by making previously documented insights, methodologies, and findings readily discoverable. Such a system can enable asynchronous collaboration and help knowledge diffuse between members. This is of particular value for new team members, who otherwise might not know what relevant source material to search. 

\subsection{Special Considerations for Researchers}
A specialized RAG system for research groups offers powerful information retrieval capabilities, but adoption depends on addressing certain deployment considerations. Many research groups prefer to maintain complete control over their infrastructure and data, sometimes due to confidentiality concerns, but often simply to ensure operational independence and sustainability. By deploying single-tenant solutions on their own infrastructure, groups avoid becoming dependent on third-party services that may change pricing models, modify features, or even discontinue operations—disruptions that could significantly impact research. In order to avoid depending on outside services, research groups will often prefer to use a local NAS instead of Google Drive or Box, and use a group-maintained Jupyterhub instance instead of Google Colab. A RAG system for research groups must align with this self-hosted ethos to fit the needs and preferences of academics. The core itself must be open source, and where outside dependencies are difficult to avoid, for example, inference on high-performing LLMs, the software must not tie the users to a particular provider.

Many research groups lack dedicated IT resources to manage complex deployments. Research groups operate with limited resources and cannot allocate significant time or money to system administration tasks that divert attention from their core scientific mission. A solution that requires extensive server configuration or database maintenance will likely remain unused regardless of its theoretical benefits. A RAG tool for researchers must therefore also prioritize minimal deployment overhead, ideally offering simple deployment procedures that researchers with only basic systems administration training can follow, and the ongoing maintenance burden must be minimal to ensure long-term sustainability.  

Finally, a solution for research groups should integrate with their existing ecosystem. It should be easy for users to import documents from repositories relevant to academics, and should support single sign-on through the identity providers most commonly used by academics and universities.

\section{\aquillm}
\aquillm is an open-source collaborative RAG tool specifically designed to address the unique requirements of research groups. It is designed to capture a wide range of document, from formal publications to instant messages, with a variety of privacy levels, as illustrated in Figure \ref{fig:datatypes}.  It aligns with the self-hosted ethos preferred by academics by providing the greatest possible degree of infrastructure and data control, while being easy to deploy and maintain. Built on established, reliable technologies including Django and Postgres, it intentionally avoids dependence on emerging AI-centered libraries like Langchain, ensuring long-term stability and maintainability while offering greater flexibility for developers.

To accommodate research groups with limited IT resources, \aquillm prioritizes minimal deployment overhead. Small groups can deploy the entire system using a single bash script on most Linux devices, including on-premise hardware, instances on commercial clouds like AWS, Google Cloud, or Azure, or academic cloud platforms such as Jetstream2. \aquillm features a custom abstraction layer for various LLM APIs, which prevents vendor lock, and supports advanced capabilities like tool calling. \aquillm can integrate with Ollama, an open source tool for hosting models locally, for completely on-premise deployment with open source models when data sovereignty is paramount \cite{ollama2023}. The installation process requires only cloning the repository, populating a configuration file, and running a shell script, and updates require only a pull and a restart, ensuring that research groups can focus on their work rather than system administration.

\aquillm supports importing papers directly from arXiv and Zotero, and supports login through the most popular identity providers, including Google, Microsoft, and GitHub. 

\subsection{Using \aquillm}

Like other RAG tools, interaction with \aquillm takes two forms: ingestion and conversation. In ingestion, users upload documents and use integrations with arXiv and Zotero to get their information into \aquillm's database. In conversation, users use a chat interface to converse with an LLM with access to the database. This workflow is depicted in Figure \ref{fig:flowchart}, along with the user interfaces that \aquillm provides for these core functions.

% \begin{figure*}
%     \centering
%     \includegraphics[width=0.8\textwidth]{collection.png}
%     \caption{\aquillm's collection management view.}
%     \label{fig:collectionview}
% \end{figure*}

\aquillm allows users to create collections to hold their documents. Collections can be shared with other users, and have a robust permission system, allowing collection owners to specify which users can read and modify a shared collection. Like directories in a file system, collections can also contain other collections. In conversation, the user can specify which collections the LLM should query. The interface for managing collections, seen at bottom right in Figure \ref{fig:flowchart}, was designed to be similar to Google Drive, Box, and other online storage solutions.  

% \begin{figure*}
%     \centering
%     \includegraphics[width=0.8\textwidth]{chat.png}
%     \caption{\aquillm's chat view.}
%     \label{fig:chatview}
% \end{figure*}

Conversation in \aquillm uses a familiar chat interface, seen at bottom left in Figure \ref{fig:flowchart}. It is broadly similar to the interfaces of ChatGPT and Claude, except that it includes an interface for selecting which collections to give the LLM access to. Unlike many RAG tools, \aquillm uses tool calling to give control of search functions to the LLM.  More basic RAG tools instead simply search the vector database using the user's message as the search string, then append the returned chunks of text to the call to the LLM. The more sophisticated approach used in \aquillm allows the model to explore the collection in order to find the information it needs to answer the user's question.

\subsection{Security}

Security is a critical consideration for applications designed to handle personal documents, such as notes, and communications, such as emails. Research groups are often reluctant to use tools that give third parties access to this information. With this in mind, \aquillm is designed to allow user groups to configure their deployment to meet their specific security needs. In the most extreme case, it can be deployed on a private server behind a VPN and configured to use local models such that no group data ever leaves group hardware. In cases with less severe security demands, it can be deployed on a cloud instance and use higher performing, less hardware-intensive LLM API calls rather than local inference. User authentication is implemented with allauth, an extension to Django that adds robust single sign-on support through a variety of providers. \textit{Within} a deployment, security is achieved with a permission system. Collections of documents are private to their creator by default and view and edit permissions can be granted to other users at the discretion of collection owners. 
%\subsection{Benefits Relative to Alternatives}
%    \begin{enumerate}
%        \item Open source, compatible with various models.
%        \item Easy to deploy.
%        \item Built using classic frameworks, requires no LLM or RAG specific knowledge to deploy and maintain.
%    \end{enumerate}
\subsection{Architecture}
\begin{figure*}
    \centering
    \includegraphics[width=0.7\linewidth]{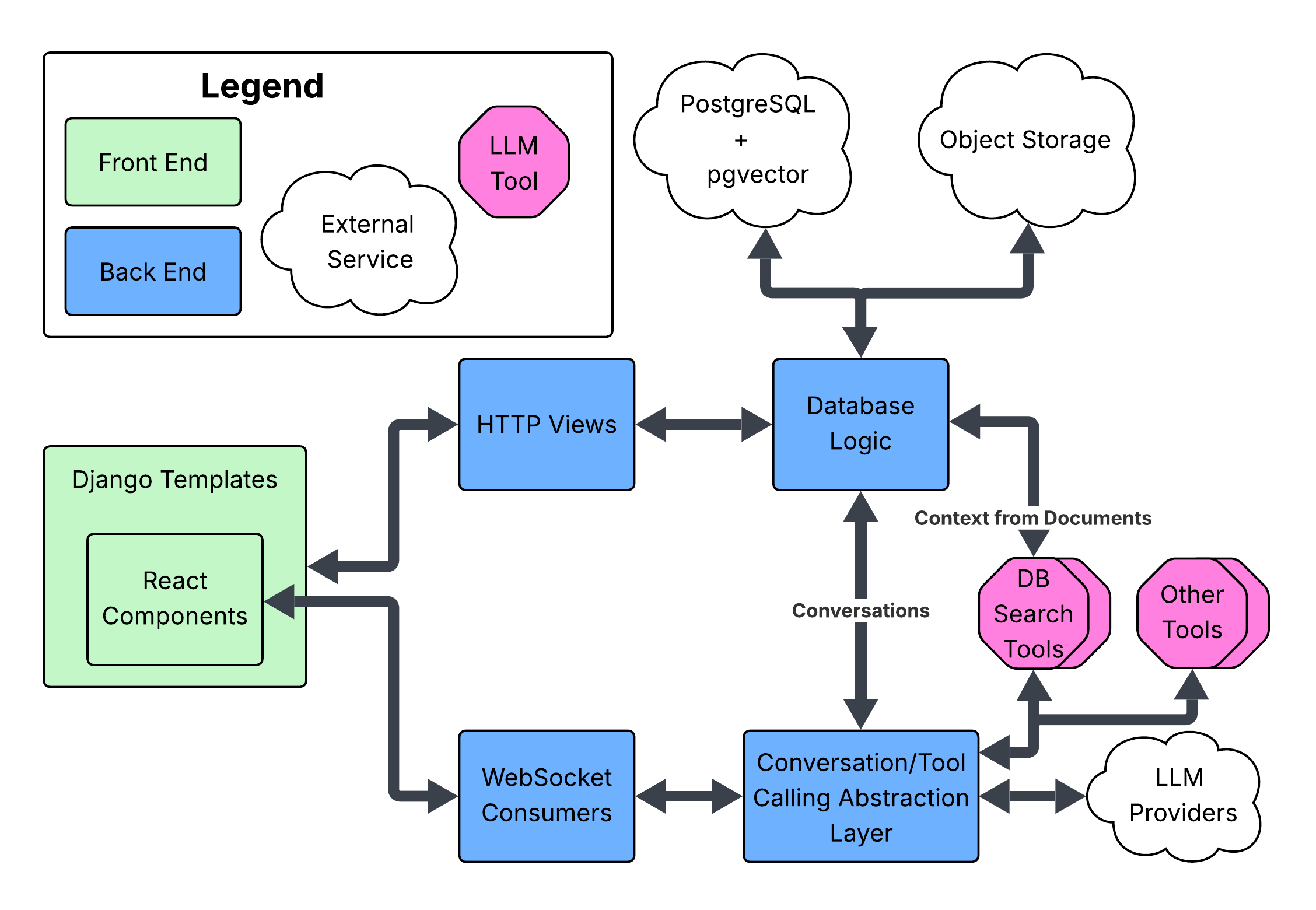}
    \caption{Architecture of \aquillm.}
    \label{fig:architecture}
\end{figure*}
\aquillm's architecture is depicted in Figure \ref{fig:architecture}. \aquillm is composed of a Django backend with custom LLM, tool calling and RAG integration logic, a front end composed of Django templates, many of which contain React components, and external services: the database, object storage, and LLM inference API. \aquillm's focus on easy deployment and maintenance for small teams drove a number of technical choices. This subsection will explore aspects of the architecture of \aquillm as they relate to design goals. From the start, the developers of \aquillm took the perspective that AI apps are fundamentally similar to other apps, but with a relatively small number of AI-related operations, such as semantic search via embeddings and LLM inference added to the mix, and that \aquillm should be thought of first as a web app and then as an AI app. In line with this philosophy, \aquillm avoids AI-specific technologies wherever possible and favors battle-tested web development technologies, including Django and PostgreSQL. These were chosen for reliability, stability, and to minimize the domain-specific knowledge required for contributors to \aquillm and maintainers of \aquillm deployments. Notably, \aquillm totally eschews LLM integration frameworks in favor of custom-written LLM-specific business logic. In addition to the flexibility this provides in defining behavior, this choice allows \aquillm to avoid a heavy dependency that could present a maintenance burden to groups using \aquillm. 

Django is a strong choice for small teams building single-tenant applications such as \aquillm due to its monolithic architecture and ``batteries included'' philosophy, providing essential components like authentication, ORM, and admin in one framework, traits which enabled the developers of \aquillm to build a secure application with complex business logic in a short amount of time. Crucially, through django-allauth, Django apps can easily support single sign-on via a huge number of social identity providers, many of which are used by universities and academics, including Google, Microsoft, Okta, ORCID, Globus, GitHub and GitLab. Because Django is a monolith, it can be difficult to scale out to large numbers of users, but because \aquillm is designed for single-tenant deployments with at most hundreds of users, this does not pose a problem. 

\subsubsection{Database and Object Storage}
As a vector search-based retrieval-augmented generation application, \aquillm requires a vector database to store text embedding vectors and retrieve them based on Euclidean distance or cosine similarity to the embedding of a search string. The last several years have seen an explosion of new vector databases, most of them developed by start-ups, including Chroma, Milvus and Pinecone. Including an additional database as a dependency to an application must be carefully considered. It raises important considerations regarding deployment, backups, migration to new versions, and long-term support that resource-constrained teams would rather avoid. With these considerations in mind, the \aquillm team elected to use pgvector, an extension to PostgreSQL that adds support for vectors and vector search indices. With the exception of object storage, \aquillm deployments keep all of their data in a single Postgres database. \aquillm's deployment scripts deploy Postgres with pgvector in a Docker container. Should users desire a fully managed solution, AWS, Azure and Google Cloud all offer Postgres as a service, and all three support the pgvector extension. 

In order to act as a central hub for team knowledge, \aquillm stores a variety of user files, including PDFs and audio. To manage these files, \aquillm uses the AWS S3 API, which is supported by a variety of locally hosted object storage tools, including LocalStack and MinIO. \aquillm's packaged install scripts deploy MinIO in Docker. Again, if user groups desire a fully managed solution, S3-compatible object storage is available from AWS and Google Cloud. 

\subsubsection{LLM Abstraction Layer, Tool Calling and RAG Logic}

Many other RAG-LLM tools use LLM integration frameworks such as Langchain or Haystack to manage AI-related parts of the application, such as calls to LLMs, tool calling, and management of vector databases. These are suitable for some applications, and are especially useful for proof of concept work, but come with significant drawbacks. First, adding a dependency which must be kept up to date adds maintenance overhead for developers. Second, as with any library, LLM integration frameworks bind the application to a particular way of doing things chosen by the author of the library. For these reasons, \aquillm eschews an LLM integration framework in favor of custom-written logic.
\paragraph{LLM Abstraction Layer}

\begin{figure*}[h]
    \includegraphics[width=\textwidth]{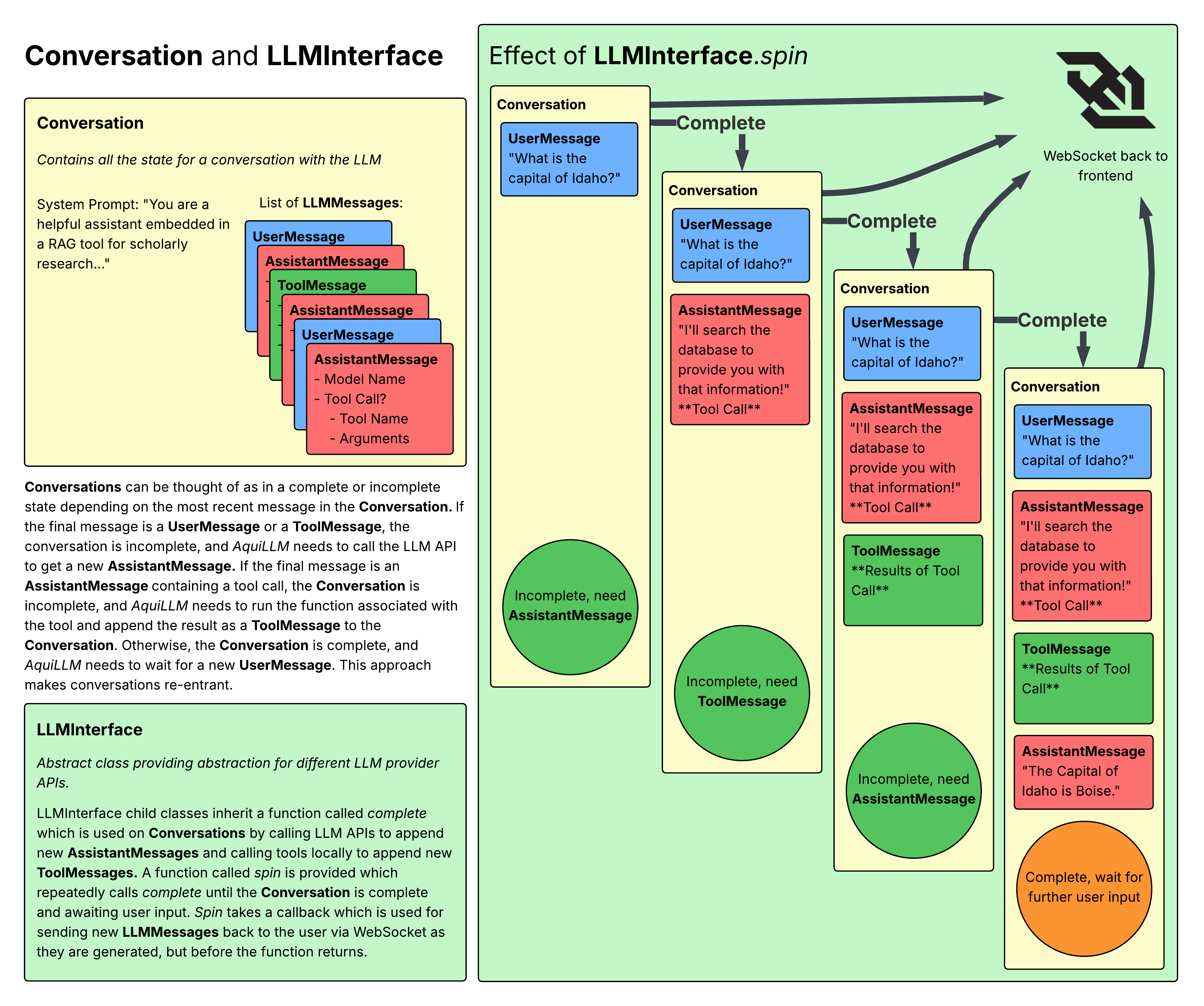}
    \caption{The core classes of \aquillm's LLM conversation management layer.}
    \label{fig:llminterface}
\end{figure*}

\aquillm features a simple yet robust LLM abstraction layer, which fulfills the following functions: compatibility between LLM providers, conversation management, and tool calling. This is accomplished by three core classes: \texttt{Conversation}, \texttt{LLMInterface}, from which interfaces for particular LLM APIs inherit, and \texttt{LLMTool}, along with a decorator for transforming ordinary Python functions into instances of \texttt{LLMTool}. The relationship between these components is illustrated in Figure \ref{fig:llminterface}. The tool calling interface uses a combination of Python's introspection facilities and Pydantic to automatically generate tool call schema in the formats LLMs expect, and validate LLM tool call arguments. To make a Python function an LLM tool, the user need only apply a decorator to the function. 

\paragraph{RAG Logic}

\aquillm's RAG logic is achieved entirely through ordinary Django ORM models. The application features models for \texttt{Collections}, which are groups of documents, \texttt{Document}, which is an abstract class from which \texttt{PDFDocument}, \texttt{TeXDocument} and other specialized children inherit, and \texttt{TextChunk}, which are the document chunks whose vector embeddings actually appear in the vector index. Chunking and embedding logic is handled by overriding the basic \texttt{save} method for \texttt{Document}, and consistency concerns specific to RAG apps (chunks should not outlive the document they come from, the same document should not appear in the same collection twice, etc.) are handled by ordinary database constraints. \texttt{TextChunk} features a class method for hybrid search (vector plus trigram) with reranking. The WebSocket consumer responsible for the front end chat interface then provides \texttt{TextChunk}'s search functionality to the LLM using the \texttt{LLMTool} decorator, completing the RAG pipeline.
\section{Progress and Preliminary Findings}

At time of writing, we have a fully-functional beta version of \aquillm which is deployed for a group of astronomers at UCLA. It is deployed to a Jetstream2 instance using the packaged deployment scripts, and accessed over the public internet \cite{jetstream}. Users have ingested papers relevant to their areas of research, as well as meeting notes and transcripts. A user who joined the lab recently has used \aquillm to `catch up' on the group's research and understand why certain decisions were made by other members before their arrival (recall that this is one of the specific problems that \aquillm was designed to address) and found its performance on this task to be useful. Other users have found it to improve querying documentation related to datasets and instruments. 

A second beta user group, environmental scientists involved in processing images from wildlife camera traps, is in the process of collecting recordings of meetings and training sessions. This information will be ingested into \aquillm to investigate its utility on such informal data. 
\section{Conclusion}
\aquillm addresses a key gap in research infrastructure by providing a lightweight, modular system for accessing the often-overlooked tacit knowledge embedded in informal and private team resources. Its architecture supports multiple document types, configurable privacy layers, and seamless integration with existing knowledge bases. By focusing on the practical needs of scholarly research groups—rather than general-purpose document retrieval—AquiLLM enables more effective collaboration, onboarding, and continuity within teams. Future work will expand its multimodal capabilities and evaluate its use across diverse scientific domains, with the broader goal of supporting sustainable, knowledge-rich research environments.

\section*{Acknowledgements}

\aquillm is funded by the Sloan Foundation, grant G-2024-22740. We would like to thank contributing Southern Oregon University students Skyler Acosta, Kevin Donlon, Elyjah Kiehne, and Jacob Nowack, as well as the members of the Astrophysics Data Lab at University of California, Los Angeles.

\bibliographystyle{plain}

\bibliography{aquillmrefs}
%% should have acknowledgement here for sloan foundation and grant # G-2024-22720

\end{document}